\documentclass[nofootinbib,aps,11pt]{revtex4}
\usepackage{slashed}
\usepackage{graphicx}% Include figure files
\usepackage{epsfig}
\usepackage{subfigure}
\usepackage{dcolumn}% Align table columns on decimal point
\usepackage{bm}% bold math
\usepackage{cancel}
\def\gsim{\lower0.5ex\hbox{$\:\buildrel >\over\sim\:$}}
\def\lsim{\lower0.5ex\hbox{$\:\buildrel <\over\sim\:$}}

\begin{document}

\title{\Large Neutrino Masses, Dark Matter and B-L Symmetry at the LHC}

\bigskip
\author{Tong Li$^{1,2}$\footnote{Electronic address: nklitong@hotmail.com}}
\author{Wei Chao$^2$\footnote{Electronic address: chaow@pku.edu.cn}}
%\footnote{email: nklitong@hotmail.com, communication
%author.}}
\address{
$^1$Bartol Research Institute, Department of Physics and Astronomy, University of Delaware, Newark, DE 19716, USA\\
$^2$Center for High Energy Physics, Peking University, Beijing 100871,
P.R.~China}
\date{\today}

%\author{Wei Chao}
%\email{chaow@pku.edu.cn}
%\author{Tong Li}
%\email{nklitong@hotmail.com}

%\address{Center for High Energy Physics, Peking University, Beijing
%100871, P.R. China \vspace{2.5cm} }

\begin{abstract}
We establish a hybrid seesaw mechanism to explain small neutrino
masses and predict cold dark matter candidate in the context of
the $B-L$ gauge symmetry extension of the Standard Model. In this model
a new scalar doublet and two new fermion singlets are introduced at
loop-level beyond the minimal Type I seesaw. The lightest particle
inside the loop can be dark matter candidate. We study in detail the
constraints from neutrino oscillation data, lepton flavor violating
processes and cosmological observation. We also explore the
predictions of the decays of the new charged scalars in each spectrum of
neutrino masses and show the most optimistic scenarios to distinguish
the spectra. We consider the pair production of the stable fermion
associated with two observable SM charged leptons at the LHC, which occurs in a 
two-step cascade decay of the new gauge boson $Z'$ and the new charged
scalars stand as intermediate particles. The masses of missing
dark matter and its parent particle can be well-determined in such
production topology.
\end{abstract}
\maketitle

%%%%%%%%%%%%%%%%%%%%%%%
\section{Introduction}
%%%%%%%%%%%%%%%%%%%%%%%
The observation of neutrino oscillations~\cite{sno,abcc,kamland,k2k}
has revealed that neutrinos have small but non-zero masses, which
cannot be accommodated in the Standard Model (SM) without
introducing extra components. Besides, the SM does not provide any candidate for the
non-baryonic cold dark matter (DM), which has been
confirmed by precisely cosmological observations~\cite{wimp}
\begin{eqnarray}
\Omega_D h^2 =0.1123 \pm 0.0035\; .
\end{eqnarray}
There are also some possible DM events from direct search setting
upper-limit on the DM-nucleon scattering
cross-section~\cite{dd1,dd2,dd3,dd4,dd5}, although they still cannot
be interpreted as significant evidence. To accommodate these
observations, the minimal SM must be extended. As so, the dark matter
and neutrino physics offer an ideal window to search for physics
beyond SM.

It is well known that neutrinos can be Dirac or Majorana fermions.
Up to now, the most attractive mechanism responsible for the origin
of small neutrino masses is the so-called seesaw mechanism, by
which one can generate Majorana masses of three active neutrinos
with renormalizable operators at the tree level. There are three
different ways to realize the tree level seesaw mechanism, categorized
as Type I~\cite{seesawI}, Type II~\cite{seesawII} and Type
III~\cite{seesawIII} seesaw mechanisms, which can all be naturally
embedded into more fundamental frameworks such as the grand unified
theories (GUTs) or string theory. The simplest and well-studied
mechanism for generating neutrino masses is the Type I seesaw, in which at
least two right-handed neutrinos are introduced. In this model,
experimental light neutrino masses are explained by the ratio of electroweak scale to large
Majorana mass term $(M_N)$ which is a suppression factor, given by $Y_\nu^2v_0^2/M_N$. The
non-ambiguous test of the nature of Majorana neutrinos, and thus a
possible test of the seesaw mechanism, will be the observation of
the lepton number violating (LNV) processes. Since the CERN Large
Hadron Collider (LHC) is going to lead us to a new energy frontier,
searching for the heavy Majorana neutrinos at the LHC appears to be
very appealing. The ideal production channel to study the LNV
processes of heavy Majorana neutrino at hadron collider is the
Drell-Yan process via SM gauge boson, $pp\to W^\ast\to \ell N$.
However, due to the rather small mixing between the heavy neutrinos
and the SM leptons in minimal Type I seesaw, to the order
$\mathcal{O}(m_\nu/M_N)$, the predicted effects of lepton number
violation are unlikely to be observable.

The possibility of testing the heavy Majorana neutrino at the LHC
beyond the minimal Type I seesaw has been considered by several groups.
Recently one group (including one of us) explored the $U(1)_{B-L}$
extension of SM~\cite{ourbl}. It is well known that $B-L$ is an
accidental global symmetry in the SM and its origin is unknown. In
order to understand the origin of Majorana neutrino masses it is
crucial to look for new scenarios where $B-L$ is a local symmetry
and can be spontaneously broken. The inclusion of three right-handed
neutrinos provides an anomaly-free formulation for the gauged
$U(1)_{B-L}$. The new gauge boson $Z_{B-L}$ can be produced at the
LHC via its gauge interactions with the quarks. Its subsequent decay
into a pair of heavy Majorana neutrinos leads to a large sample of
events without involving the small mixing suppression. The $\Delta
L=2$ signal $pp\to Z'\to NN\to \ell^\pm\ell^\pm W^\mp W^\mp$ with
two $W$ bosons hadronic decay will directly test the nature of heavy Majorana
neutrinos. However, because no stringent experiments constrain the mass
of heavy Majorana neutrino up to now, a pessimistic case with heavy
Majorana neutrino mass larger than half of $Z_{B-L}$ boson mass
$M_N>M_{Z'}/2$ may likely happen in the reality. Consequently one
cannot see any lepton number violating signatures at hadron
collider because they are kinematically forbidden. This dangerous
situation motivates us to figure out other neutrino mass generation
mechanisms which provide observable signatures in the $U(1)_{B-L}$ extension
framework.

On the other hand, if one introduces only two types of fields,
scalar $\eta$ and/or fermion $\psi$, neutrino masses can be
generated at loop level~\cite{zee,babu,ma,threeloop}. A salient
feature of such radiative seesaw mechanisms is that all the new
particles can lie below several TeVs naturally and their
interactions with SM particles are not heavily constrained by
electroweak precision measurements so that they can be directly
tested at the LHC. Other discrete symmetries, for instance $Z_2$
symmetry, can be imposed by hand to forbid tree level neutrino mass
term if new scalar $\eta$ and fermion $\psi$ stay in the same
representation as the scalar doublet in SM or the scalar triplet
$\Delta\sim (1,3,1)$ in Type II seesaw, fermionic singlets
$N\sim(1,1,0)$ in Type I seesaw or fermionic triplets
$\Sigma\sim(1,3,0)$ in Type III seesaw respectively. Consequently
the new particles inside the loop only self-interact and couple to
SM leptons, and the lightest one is stable which can serve as cold
dark matter candidate. This feature makes it possible to relate the
dark matter with neutrino physics. If we restrict our attention to
radiative seesaw extended by an extra Abelian gauge symmetry or new
representation of other non-Abelian gauge groups, for instance
$U(1)_{B-L}$ extension or adjoint representation of
$SU(3)$~\cite{wise}, such feature can also be achieved.

In this paper we investigate a particular realization of a radiative
seesaw model in the $U(1)_{B-L}$ extension of SM. In the exact
$U(1)_{B-L}$ extension of SM, three right-handed neutrinos $N_i$ are
introduced and a new scalar field $\Phi$ is added to break the local
$B-L$ symmetry. Besides this framework, we add an extra scalar
doublet $\eta^T=(\eta^+, \eta^0)$ and two fermions $\psi_i$ in which
$\eta$ has no vacuum expectation value. The $B-L$ charges of $\eta$,
$\psi_i$ and $\Phi$ are $+1$, $0$ and $+2$, respectively. In this
model, there are two terms contributing to neutrino masses: tree
level term coming from the modified Type I seesaw mechanism with
right-handed neutrino masses generated after $U(1)_{B-L}$ symmetry
breaking, and one-loop level term mediated by the new scalar $\eta$ and
fermion $\psi_i$. The constraints from lepton flavor violating
processes are also studied. We focus on $m_\psi<m_\eta$ scenario in
which the lighter $\psi$ could serve as the cold dark matter candidate.
Relevant cosmological constraints are investigated. Due to the
existence of $U(1)_{B-L}$ gauge symmetry, charged scalar $\eta^\pm$
and dark matter $\psi$ have clear signatures associated with SM
charged leptons at the LHC: $pp\to Z'\to \eta^+\eta^-\to
\ell^+\ell^-\psi\psi$, even if heavy Majorana neutrinos are
forbidden to be produced. We find encouraging results for the LHC
signatures of this model to learn about the light neutrino
properties and to well-determine the masses of dark matter $\psi$
and its parent particle $\eta$ in such a production topology.

This work is organized as follows: In Section II, we describe our
model for neutrino masses and dark matter. In Section III, we
discuss the constraints on the neutrino mass and mixing parameters
from the current neutrino oscillation data, lepton flavor violating
processes and cosmological observation. Section IV is devoted to
discuss the possibility of detecting the model at the LHC. We
summarize our findings in Section V. Feynman rules for new particles
are presented in the Appendix.

\section{The Neutrino Mass and Dark Matter Model}

In our model the SM is extended with three right-handed Majorana
neutrinos $N_i\sim (1,1,0)$, one scalar singlet $\Phi\sim (1,1,0)$,
one scalar doublet $\eta\sim (1,2,1/2)$ and two right-handed
Majorana fermions $\psi_i\sim (1,1,0)$, as well as $U(1)_{B-L}$
gauge symmetry. The $B-L$ charges for all fields are collected in
Table.~\ref{BL}.
\begin{table}[tb]
\begin{center}
\begin{tabular}[t]{|c|c|c|c|c|c|c|c|c|}
  \hline
   & $Q_L,u_R,d_R$ & $l_L,\ell_R$ & $N_R$ & $H$ & $\Phi$ && $\psi$ & $\eta$\\
  \hline
  $B-L$ & ${1\over 3}$ & $-1$ & $-1$ & $0$ & $+2$ && $0$ & $+1$\\
\hline
 \end{tabular}
\end{center}
\caption{Fields and their $B-L$ charges in our model, where
$\ell=e,\mu,\tau$. The fields on the left-handed side of the
double-vertical lines are the exact contents of minimal local
$U(1)_{B-L}$ extension of SM.} \label{BL}
\end{table}

With well-chosen charges, as a result, the new lagrangian can be
written as
\begin{eqnarray}
{\cal L}={\cal L}_{Kinetic}+{\cal L}_{Scalar}-\left[Y_\psi^{}
\overline{l_L^{}}\widetilde{\eta}\psi_R^{} + Y_\nu^{}
\overline{l_L^{}} \widetilde H N_R^{} +{1\over 2} m_\psi^{}
\overline{\psi^C_R}\psi_R^{} + {1\over 2} Y_N^{}
\overline{N^C_R}N_R^{}  \Phi+ {\rm h.c.} \right]\label{lagrangian}
\end{eqnarray}
where $l_L^{} =(\nu_L^{}, \ell_L^{})$ and $H^T=(H^+,H^0)$ are the
left-handed lepton doublet and scalar doublet in SM respectively
with $\widetilde{H}(\widetilde{\eta})=i\sigma_2H^\ast(\eta^\ast)$.
The kinetic term for the matter fields and Lagrangian for scalar
fields are
\begin{eqnarray}
\mathcal{L}_{Kinetic}&=&i\overline{Q_L}\gamma^\mu D_\mu
Q_L+i\overline{u_R}\gamma^\mu D_\mu u_R+i\overline{d_R}\gamma^\mu
D_\mu d_R+i\overline{l_L}\gamma^\mu D_\mu
l_L+i\overline{\ell_R}\gamma^\mu D_\mu
\ell_R+i\overline{N_R}\gamma^\mu D_\mu N_R\nonumber \\
&+&i\overline{\psi_R}\gamma^\mu D_\mu \psi_R,\\
\mathcal{L}_{Scalar}&=&(D_\mu H)^\dagger (D^\mu H)+(D_\mu
\eta)^\dagger (D^\mu \eta)+(D_\mu \Phi)^\dagger (D^\mu
\Phi)-V(H,\eta,\Phi)
\end{eqnarray}
where
\begin{eqnarray}
D_\mu N_R=\partial_\mu N_R-ig_{BL}B'_\mu N_R , \ D_\mu
\psi_R=\partial_\mu \psi_R, \ D_\mu \Phi=\partial_\mu
\Phi+i2g_{BL}B'_\mu \Phi.
\end{eqnarray}
Here $B'_\mu$ and $g_{BL}$ are the gauge field and gauge coupling
for $U(1)_{B-L}$ respectively. The Higgs potential is given by
\begin{eqnarray}
V(H,\eta,\Phi)&=&-m_H^2H^\dagger H -m_\eta^2\eta^\dagger \eta
-m_\Phi^2\Phi^\dagger \Phi +\lambda_H(H^\dagger H)^2+\lambda_\eta
(\eta^\dagger \eta)^2+\lambda_\Phi (\Phi^\dagger
\Phi)^2\nonumber \\
&+&\lambda_1^{} (H^\dagger H)(\eta^\dagger \eta)+\lambda_2^{}
(H^\dagger \eta)(\eta^\dagger H)+\lambda_3^{} (H^\dagger
H)(\Phi^\dagger \Phi)
+\lambda_4^{} (\eta^\dagger \eta)(\Phi^\dagger \Phi)\nonumber \\
&+&{\lambda_5^{} \over \Lambda } \left[(H\eta^\dagger)^2\Phi+ h.c.
\right ]\label{potential}
\end{eqnarray}
where $\lambda_5^{}$ has been set to be real without losing generality
and $\Lambda$ is a new high energy scale. The last term is a
dimension-5 effective operator in this model which has an accidental
$B-L$ symmetry. Because of $B-L$ gauge invariance, new fermions
$\psi_i$ only couple to the new scalar doublet $\eta$ and
right-handed neutrinos only couple to SM Higgs $H$. For $m_H^{2},
m_\Phi^2
>0$ and $m_\eta^2<0$, only $H$ and $\Phi$ acquire nonzero vacuum
expectation values (VEVs). The $U(1)_{B-L}^{}$ symmetry is
spontaneously broken by VEV of $\Phi$. After imposing the conditions
of global minimum, one obtains
\begin{eqnarray}
v_0^2 &=& {\lambda_\Phi^{} m_H^2 - \lambda_3^{} m_\Phi^2 \over
\lambda_H^{} \lambda_\Phi^{} -\lambda_3^{2}} \; ,  \\
v_\Phi^2 &=& {\lambda_H^{} m_\Phi^2 - \lambda_3^{} m_H^2 \over
\lambda_H^{} \lambda_\Phi^{} -\lambda_3^2} \; .
\end{eqnarray}
We define $H^0= (h^0+v_0+iG^0)/\sqrt{2}$, $
\eta^0=(\delta^0+iF^0)/\sqrt{2}$, $\Phi = (\phi^0+v_\Phi+iK^0)/
\sqrt{2}$, $H^\pm = G^\pm$ and $\eta^\pm=\delta^\pm$\footnote{We
denote the new scalar doublet as $\delta$ below.}, and the mass
eigenvalues of the resulting physical bosons are given by
\begin{itemize}
\item CP-even states:
\begin{eqnarray}
&&h^0, \ m_{h^0}^2=2\lambda_H v_0^2;\\
&&\phi^0, \ m_{\phi^0}^2=2\lambda_\Phi v_\Phi^2;\\
&&\delta^0, \ m_{\delta^0}^2=-m_\eta^2+{1\over
2}(\lambda_1+\lambda_2)v_0^2+{1\over 2}\lambda_4v_\Phi^2 +
{v_0^2v_\Phi\lambda_5\over \sqrt{2}\Lambda}
\end{eqnarray}
\item CP-odd state:
\begin{eqnarray}
F^0, \ m_{F^0}^2=-m_\eta^2+{1\over
2}(\lambda_1+\lambda_2)v_0^2+{1\over 2}\lambda_4v_\Phi^2 -
{v_0^2v_\Phi\lambda_5\over \sqrt{2}\Lambda}
\end{eqnarray}
\item charged state:
\begin{eqnarray}
\delta^\pm, \ m_{\delta^\pm}^2=-m_\eta^2+{1\over
2}\lambda_1v_0^2+{1\over 2}\lambda_4v_\Phi^2
\end{eqnarray}
\item new gauge boson:
\begin{eqnarray}
Z'=Z_{B-L}, \ M_{Z'}=2g_{BL}v_\Phi\label{zblmass}
\end{eqnarray}
\end{itemize}
The mixing angle between $\phi^0$ and $h^0$ is $\tan 2 \theta =
\lambda_3^{}v_0^{} v_\Phi^{}/(\lambda_H^{} v_0^2 -\lambda_\Phi^{}
v_\Phi^2)$. Relevant interactions are collected in the Appendix. For
simplicity we demand the mass hierarchy in this model as below
\begin{eqnarray}
&&m_{\psi}\equiv m_{\psi_1}\lesssim m_{\psi_2}<m_{\delta^\pm}\ll
m_{\phi^0}\sim
M_{N}\sim M_{Z'}\sim v_\Phi,\nonumber\\
&&m_\delta\equiv m_{\delta^\pm}\approx m_{\delta^0}\approx m_{F^0}.
\label{hierarchy}
\end{eqnarray}

\begin{figure}[tb]
\begin{center}
\begin{tabular}{cc}
\includegraphics[scale=1,width=9cm]{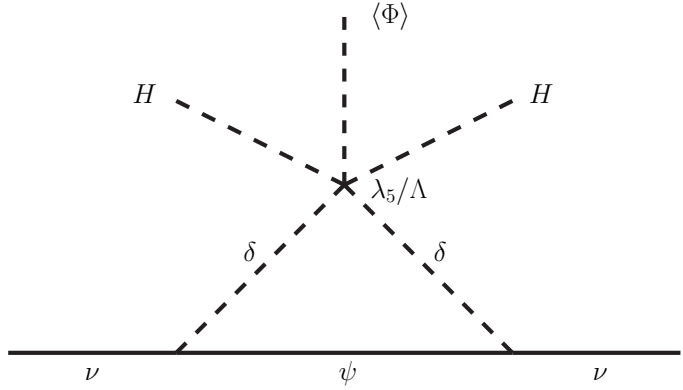}
\end{tabular}
\end{center}
\caption{Feynman diagram for loop-level neutrino mass generation in
this model.} \label{ourradiative}
\end{figure}

After $\Phi$ is set at its VEV, right-handed Majorana neutrinos acquire
masses, $M_N^{}= Y_N^{} v_\Phi^{}/\sqrt{2} $. According to
Eq.~(\ref{lagrangian}), there are two terms contributing to light
Majorana neutrino mass matrix: tree level Type I seesaw terms
$Y_\nu^{} \overline{l_L^{}} \widetilde H N_R^{},{1\over 2} Y_N^{}
\overline{N^C_R}N_R^{}  \Phi$ and one-loop radiative corrections as
depicted in Fig.~\ref{ourradiative}. Notice that
Fig.~\ref{ourradiative} is induced by the Yukawa coupling among
$\psi_i$, $\delta$ and SM leptons and the existence of
$\lambda_5^{}$ term in Eq.~(\ref{potential}). If $\lambda_5$ is
zero, neutrino masses would purely come from Type I seesaw term. In
the most general case, active neutrino mass matrix can be written as
\begin{eqnarray}
\left( M_\nu^{} \right)_{\alpha \beta}^{} = (M_{tree})_{\alpha
\beta}^{} + (M_{loop}^{} )_{\alpha \beta}^{}= -\left(M_D^{} M_N^{-1}
M_D^T \right)_{\alpha \beta}^{} + \sum_{i=1,2}^{} \left
(Y_\psi^{}\right)_{\alpha i } \left( Y_\psi^{}\right)_{\beta i}^{} {
I(m_{\psi_i}^2/m_\delta^2 ) \over m_{\psi_i}^{} } \label{hybrid}
\end{eqnarray}
where $M_D^{} = Y_\nu^{}  v_0^{}/\sqrt{2} $ is the Dirac neutrino
mass matrix and
\begin{eqnarray}
I(x)={\lambda_5 v_\Phi v_0^2\over 8\sqrt{2}\pi^2\Lambda
}\left({x\over 1-x}\right)\left[1+{x{\rm ln}x\over 1-x}\right] \; .
\end{eqnarray}
Neutrino mass formula in Eq.~(\ref{hybrid}) is called as the ``hybrid''
seesaw mechanism. There are three scenarios: (1) ${\cal O}
(M_{tree}^{}) \gg {\cal O} (M_{loop}^{})$, Type I seesaw term
dominates the contribution to the neutrino masses; (2) ${\cal O}
(M_{tree}^{}) \sim {\cal O} (M_{loop}^{})$, both terms contribute to
neutrino masses; and (3) ${\cal O} (M_{tree}^{}) \ll {\cal O}
(M_{loop}^{})$, radiative seesaw term dominates. We can impose
$Z_2^{}$ discrete symmetry for right-handed neutrino $N_R\to -N_R$
to eliminate the Type I seesaw term, but actually the detectable
phenomenology discussed later does not depend on which scenario we
work on. Because the pure Type I seesaw scenario (1) has been
well-studied by many groups both theoretically and phenomenologically, 
we mainly concentrate our scope on the other extreme
scenario (3) in Section 3.

It is important to emphasize that the last term in
Eq.~(\ref{potential}) is not the only dimension-5 effective operator
in this model. There are other dimension-5 operators allowed by
gauge symmetry, for instance ${1\over \Lambda}(H^\dagger
H+\eta^\dagger \eta+\Phi^\dagger \Phi)\overline{\psi^C_R}\psi_R,
{1\over \Lambda}H^\dagger \eta \psi_R N_R, {1\over
\Lambda}l_L^TH\Phi N_R, {1\over \Lambda}l_Ll_L\eta\eta$. Fermion
$\psi$ can get additional but suppressed mass contribution
$(v_0^2+v_\Phi^2)/\Lambda$. ${1\over \Lambda}H^\dagger \eta \psi_R
N_R$ term contributes to right-handed neutrino mass to the order
${v_0^2/\Lambda^2}$ and ${1\over \Lambda}l_L^TH\Phi N_R$ contributes
to Dirac neutrino mass term to the order $v_\Phi/\Lambda$. ${1\over
\Lambda}l_Ll_L\eta\eta$ can give extra neutrino mass term
$(m_\psi/\Lambda) M_{loop}$ in Eq.~(\ref{hybrid}). In brief, all
other dimension-5 operators supply additional suppression factor for
neutrino mass generation or lepton flavor violation. We therefore
ignore these suppressed contributions in our analysis. The usual
neutrino seesaw mass term ${1\over \Lambda}l_Ll_LHH$ is forbidden by
$U(1)_{B-L}$ gauge symmetry.

Notice that the new particles $\psi,\delta$ inside the loop only couple
to SM leptons and do not have mass mixing terms like $M_D$.
Therefore another emergent consequence of this model is the
appearance of a lightest stable particle, which can serve as the
cold dark matter. It can be bosonic~\cite{shfang} (i.e. the lighter
one of $\delta^0$ and $F^0$) or
fermionic~\cite{Ma,Suematsu,reconciliation} (i.e. the lighter one of
$\psi_i^{}$). In this paper, as shown in Eq.~(\ref{hierarchy}), we
assume $m_\psi<m_\delta$ and two fermions $\psi_1^{}$, $ \psi_2^{}$
are quasi-degenerate. Consequently the lighter $\psi$ would be the cold
dark matter candidate.

\section{Constraint on the parameter space}

In this section, we discuss the constraints on the parameter space
of the model from (a) neutrino physics, (b) lepton flavor violating
processes and (c) cosmological observation.

\subsection{Neutrino Masses from Radiative Seesaw}

We assume that there is only $M_{loop}^{}$ term left in
Eq.~(\ref{hybrid}). The neutrino mass matrix can be diagonalized as
\begin{eqnarray}
\hat M = V^\dagger_{PMNS} M_\nu^{} V^*_{PMNS}
\end{eqnarray}
where $\hat M ={diag} (m_1^{} , m_2^{}, m_3^{})$. $V_{PMNS}^{}$ is
the lepton mixing matrix, i.e. the Pontecorvo-Maki-Nakagawa-Sakata
(PMNS) matrix~\cite{PMNS}, which comes from the mismatch between the
diagonalization of neutrino masses and charged lepton mass matrix.
With the standard parametrization, $V_{PMNS}^{}$ can be written as
\begin{equation}
V_{PMNS}= \left(
\begin{array}{lll}
 c_{12} c_{13} & c_{13} s_{12} & e^{-\text{i$\delta $}} s_{13}
   \\
 -c_{12} s_{13} s_{23} e^{\text{i$\delta $}}-c_{23} s_{12} &
   c_{12} c_{23}-e^{\text{i$\delta $}} s_{12} s_{13} s_{23} &
   c_{13} s_{23} \\
 s_{12} s_{23}-e^{\text{i$\delta $}} c_{12} c_{23} s_{13} &
   -c_{23} s_{12} s_{13} e^{\text{i$\delta $}}-c_{12} s_{23} &
   c_{13} c_{23}
\end{array}
\right)\times \text{diag} (1,e^{i \chi}, 1)
\end{equation}
where $s_{ij}=\sin{\theta_{ij}}$, $c_{ij}=\cos{\theta_{ij}}$, $0 \le
\theta_{ij} \le \pi/2$ and $0 \le \delta,\chi \le 2\pi$. The phase
$\delta$ is the Dirac CP phase, and $\chi$ is the Majorana phase.
The experimental constraints on the neutrino masses and mixing
angles, at $2\sigma$ level are~\cite{Schwetz}
\begin{eqnarray}
7.25 \times 10^{-5} {\rm eV}^2 \  < & \Delta m_{21}^2 & < \  8.11 \times 10^{-5} {\rm eV}^2\; , \\
2.18 \times 10^{-3} {\rm eV}^2 \  < & |\Delta m_{31}^2| & < \  2.64 \times 10^{-3} {\rm eV}^2 \; , \\
                   0.27 \  < & \sin^2{\theta_{12}} & < \  0.35\; , \\
                   0.39 \  < & \sin^2{\theta_{23}} & <\  0.63\; , \\
                          & \sin^2{\theta_{13}} & <\  0.04 \; ,
\end{eqnarray}
and $\sum_{i} m_{i} < \ 1.2$ eV. Using Casas-Ibarra
parametrization~\cite{casibba}, one can find a formal solution for
the Yukawa coupling between the SM charged leptons
$(\ell=e,\mu,\tau)$ and $\psi_i (i=1,2)$
\begin{eqnarray}
\left(Y_\psi \right)_{\ell i}^{} = V_{PMNS}^{} \hat M^{1 /2} O
F^{-1/2} \; ,
\end{eqnarray}
in terms of a complex matrix which satisfies the orthogonality
condition $O^TO=OO^T=1$, where $F={diag}(I(m_{\psi_1}^2/
m_\delta^{})/m_{\psi_1},I( m_{\psi_2}^2/ m_\delta^{})/m_{\psi_2})$.
In the $3+2$ mode, one of the active neutrinos is massless and $\hat
M= {diag} ( { 0, ~\sqrt{\Delta m_{21}^2}, ~\sqrt{|\Delta m_{31}^2|}}
) $ for normal hierarchy (NH), $\hat M= {diag}( { \sqrt{|\Delta
m_{31}^2|}, ~\sqrt{\Delta m_{21}^2+ |\Delta m_{31}^2|}, 0 ~})$ for
inverted hierarchy (IH). The matrix $O$ can be written as~\cite{32}
\begin{eqnarray}
O= \left(
  \begin{array}{cc}
    0 & 0 \\
    \cos z & \sin z \\
    -\sin z & \cos z \\
  \end{array}
\right) \ \ \ {\rm for \ NH \ }, \ O= \left(
  \begin{array}{cc}
    \cos z & \sin z \\
    -\sin z & \cos z \\
        0 & 0 \\
  \end{array}
\right) \ \ \ {\rm for \ IH \ }
\end{eqnarray}
where $z=x+ iy$ with real parameters $x$ and $y$.

\begin{figure}[tb]
\begin{center}
\begin{tabular}{cc}
\includegraphics[scale=1,width=8cm]{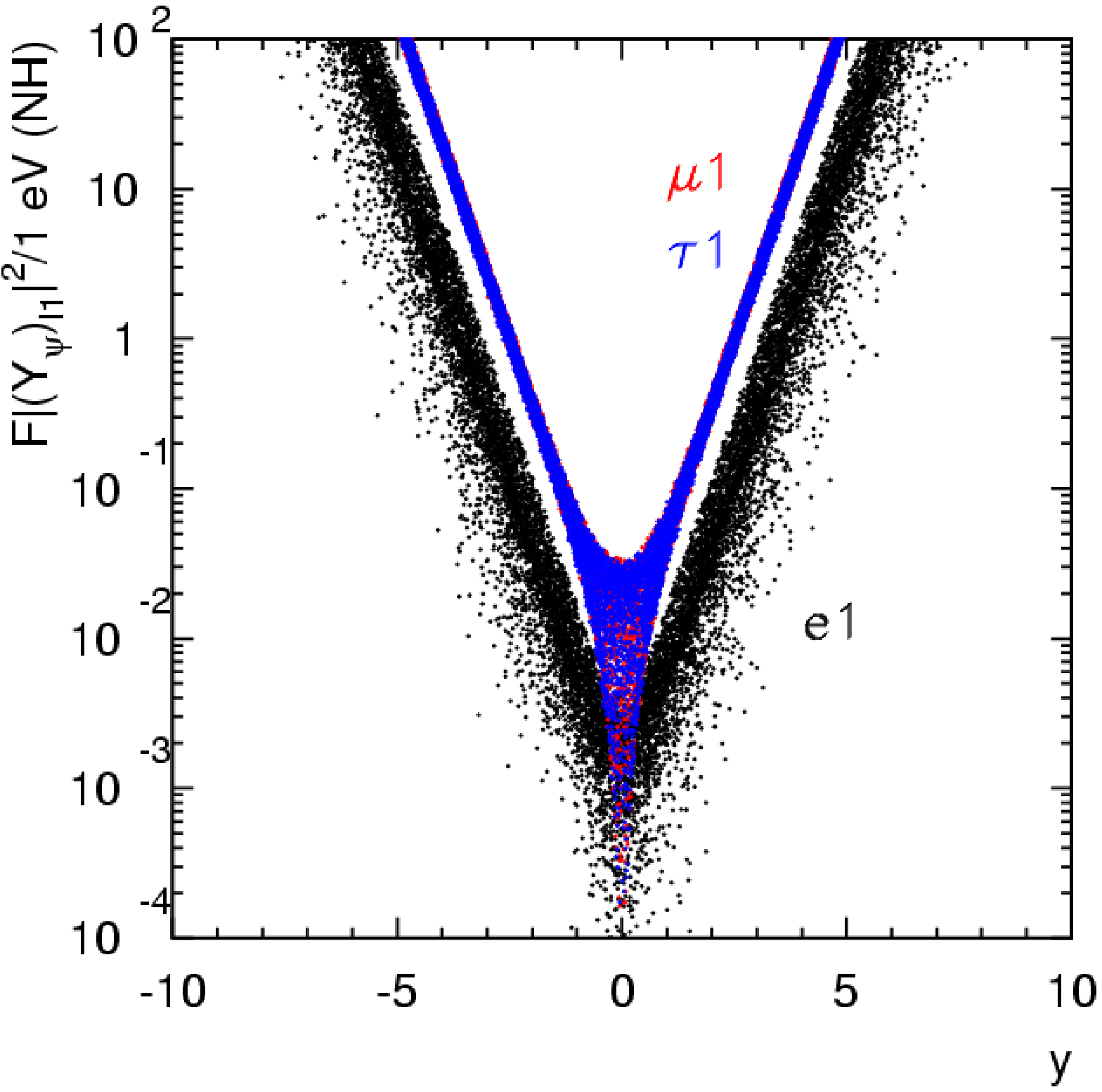}
\includegraphics[scale=1,width=8cm]{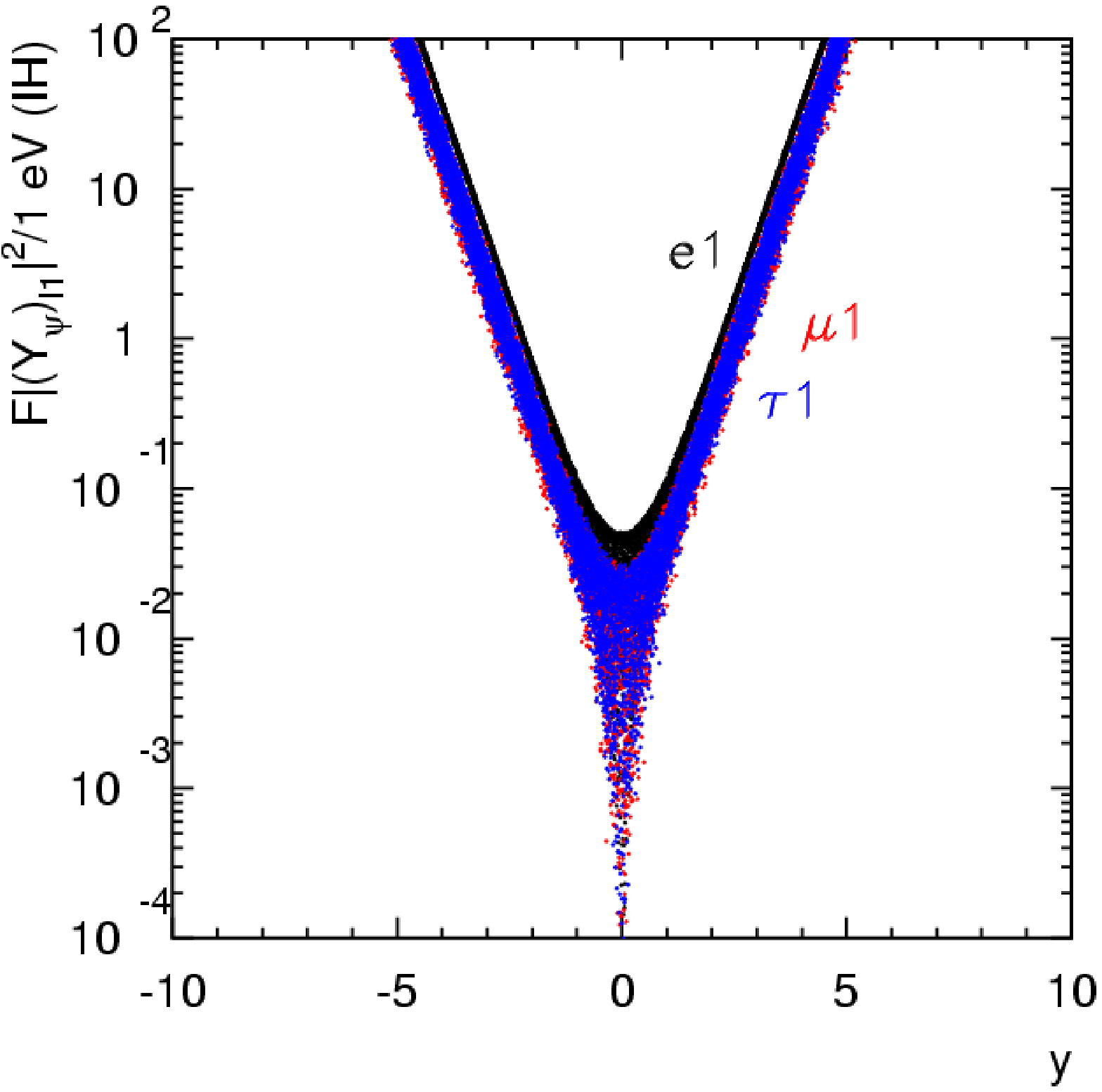}
\end{tabular}
\end{center}
\caption{$F|(Y_\psi)_{\ell 1}|^2/ 1~{\rm eV}$ versus parameter $y$
in matrix $O$ for NH (left) and IH (right), assuming vanishing
Majorana phase and $0\leq x\leq 2\pi$.} \label{vv}
\end{figure}

Assuming vanishing Majorana phase and $0\leq x\leq 2 \pi$, we show
$F |(Y_{\psi})_{\ell 1}^{}|^2/ 1 {\rm~ eV}$ $(\ell= e , \mu, \tau )$
versus parameter $y$ in Fig.~\ref{vv}. The behaviors for $F
|(Y_{\psi})_{\ell 2}^{}|^2/ 1 {\rm~ eV}$ are quite similar. One can
see $\mu,\tau$ elements are several times larger than that of
electron in NH and electron element is slightly larger in IH.

\subsection{Constraints from Lepton Flavor Violating Process}

\begin{figure}[tb]
\begin{center}
\begin{tabular}{cc}
\includegraphics[scale=1,width=8cm]{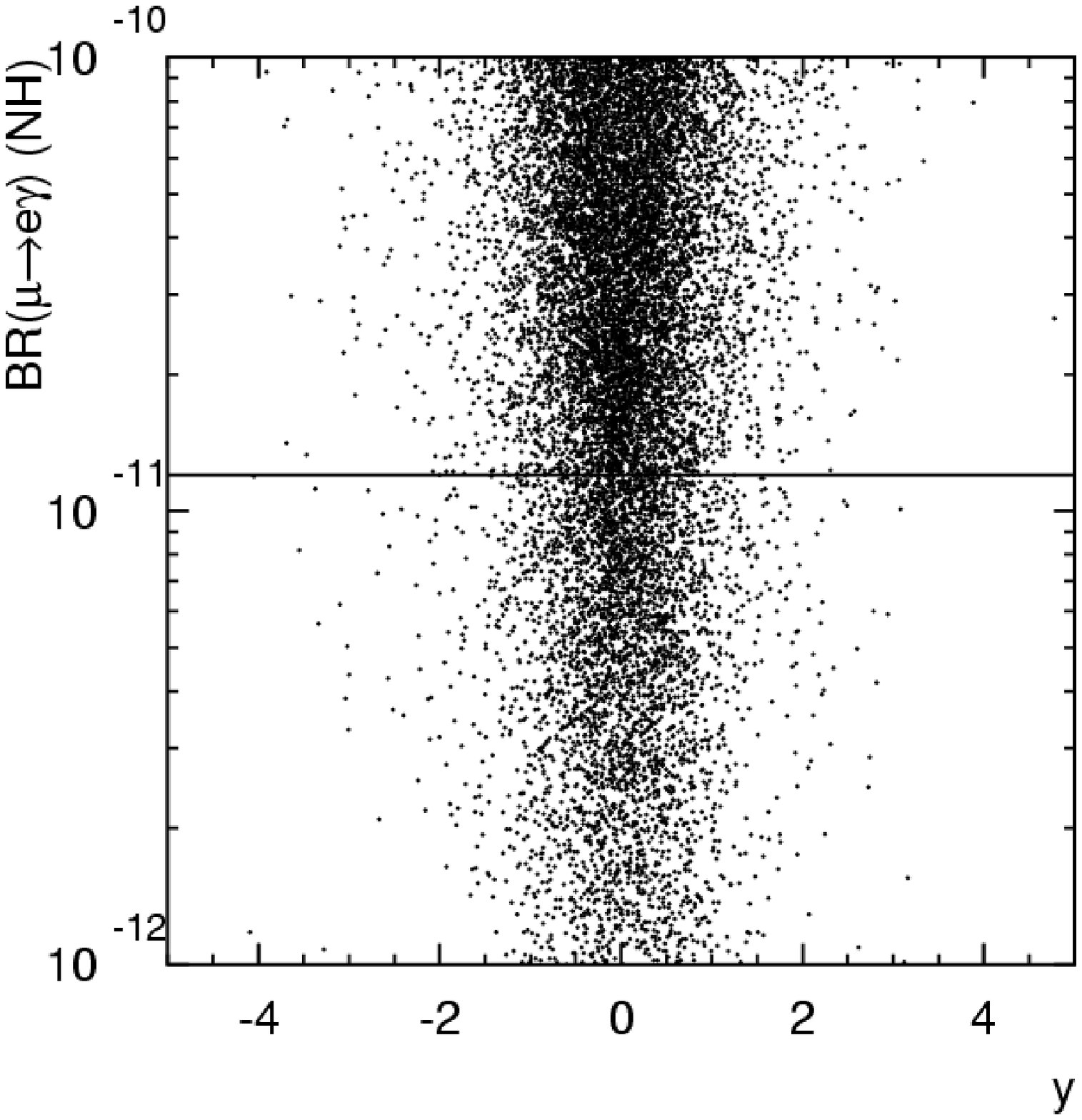}
\includegraphics[scale=1,width=8cm]{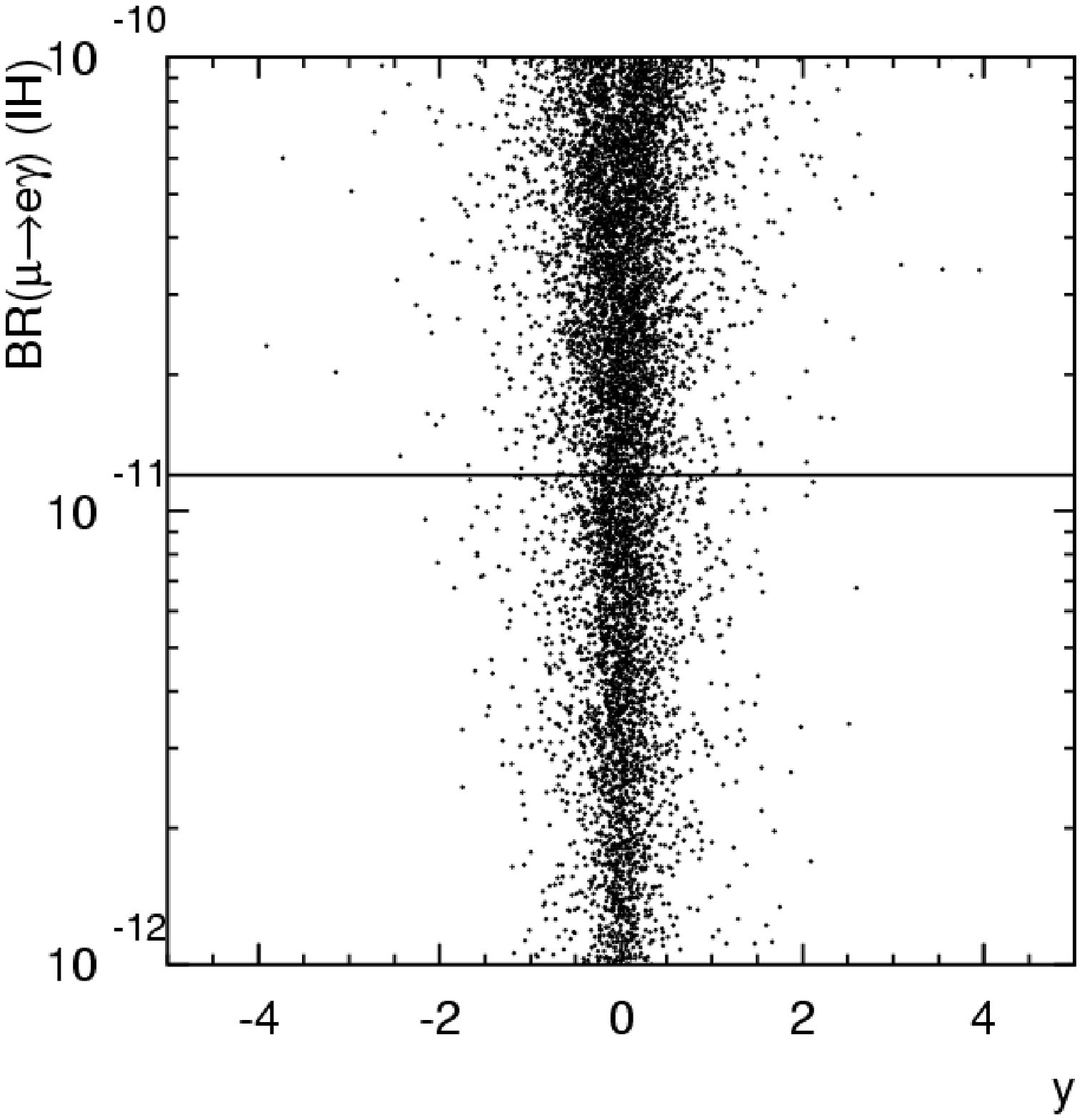}
\end{tabular}
\end{center}
\caption{The branching fraction of $\mu\to e \gamma$ versus
parameter $y$ in matrix $O$ for NH (left) and IH (right), assuming
vanishing Majorana phase, $0\leq x\leq 2\pi$, $F=1$ eV, $m_\psi=100$
GeV and $m_\delta=300$ GeV. The horizontal line represents the
current experimental bound for $BR(\mu\to e\gamma)<1.2\times
10^{-11}$.} \label{mue}
\end{figure}

Here we briefly discuss constraints on the parameter space from
lepton flavor violating (LFV) processes such as $\mu \rightarrow e
\gamma $. One-loop diagrams mediated by $\eta^\pm$ and $\psi$,
similar to the one generating small neutrino masses, contribute to
the radiative decay processes. The branching ratio of $\mu\to
e\gamma$ decay was discussed in Refs.~\cite{Ma,Suematsu,reconciliation}
\begin{eqnarray}
BR (\mu\to e\gamma)&=&{3\alpha\over 64\pi (G_Fm_\delta^2)^2}|\sum_i
(Y_{\psi})_{\mu i} (Y_\psi)_{e i}^\ast
F_2(m_{\psi_i}^2/m_\delta^2)|^2 \; ,
\end{eqnarray}
where
\begin{eqnarray}
F_2(x)&=&{1-6x+3x^2+2x^3-6x^2{\rm ln}x\over 6(1-x)^4} \; .
\end{eqnarray}
In principle, we can always tune the function of parameter
$\lambda_5$ to get appropriate Yukawa couple $Y_\psi$ to escape from
lepton flavor violating constraints. In Fig.~\ref{mue} we show the
branching fraction of $\mu \rightarrow e \gamma$ versus parameter
$y$ by assuming $0\leq x\leq 2 \pi$, $F=1~ {\rm eV}$, $m_\psi^{}
=100~ {\rm GeV}$ and $m_\delta^{} =300 ~{\rm GeV}$. The horizontal
line is the current experimental bound, i.e. $BR (\mu \rightarrow e
\gamma) < 1.2\times 10^{-11}$~\cite{pdg}. One can find that $\mu
\rightarrow e \gamma$ restricts $|y|$ to be less than $3$ for NH and
$2$ for IH. Combining the constraints from neutrino masses as shown
in Fig.~\ref{vv} with LFV in Fig.~\ref{mue}, we have Yukawa
couplings as $|(Y_\psi)_{e1}|^2\lesssim 0.5, |(Y_\psi)_{\mu 1,\tau
1}|^2\lesssim 1$ for NH and $|(Y_\psi)_{e 1, \mu 1,\tau
1}|^2\lesssim 1$ for IH, which are not stringently constrained.

\subsection{Cosmological Constraint}

Because of the mass hierarchy assumed in previous section, the
lighter one of $\psi_i^{}$ would be the dark matter candidate in this
model, denoted as $\psi$. Its relic density is calculable as a
function of Yukawa coupling $Y_{\psi}^{}$, dark matter mass
$m_{\psi}^{}$ and mass of scalar $\delta$. In this model, the
relic density of $\psi^{} $ is controlled by the annihilation of $
\psi^{} \psi^{} \rightarrow \nu \nu, ~\ell^+ \ell^-$ through the
exchange of $\delta^\pm$, $\delta^0$ and $F^0$ in t-channel. Because
the processes through $\delta^0$ and $F^0$ exchanges have huge cancellation,
two $\psi^{}$'s annihilate mainly into two charged leptons. Ignoring
charged lepton masses, one can write down the interaction rate
$\sigma v$ in non-relativistic
limit~\cite{Ma,Suematsu,reconciliation}
\begin{eqnarray}
\sigma_{ann} v_{rel}&\equiv&a+bv_{rel}^2=
\sum_{\alpha\beta}|(Y_\psi)_{\alpha 1}|^2|(Y_\psi)_{\beta
1}|^2{r_\psi^2(1-2r_\psi+2r_\psi^2)\over 24\pi m_\psi^2}v_{rel}^2 \;
,
\end{eqnarray}
where $r_\psi^{} =m_\psi^2/(m_\psi^2+m_\delta^2)$ and $v_{rel}^{}$
is relative speed. The thermally averaged cross section can be
written as $\langle \sigma_{\rm anna}^{} v_{\rm rel}^{} \rangle =a +
6 b/x_f^{}$, where $x_f^{} = m_\psi^{}/ T_f^{} $ and $T_f^{} $ is
the freeze-out temperature of the relic particle. The present
density of $\psi$ is simply given by $\rho_\psi^{}=m_\psi^{} s_0^{}
Y_\infty^{}$, where $s_0^{} = 2889.2~ {\rm cm}^{-3}$ is the present
entropy density and $Y_\infty^{}$ is the asymptotic value of the
ratio $n_\psi^{}/ s_0 $ with $ Y_\infty^{-1} = 0.264 \sqrt{g_\ast}
M_{Pl}^{} m_\psi^{} (a+3b/x_f) x_f^{-1}$ through the time (temperature) 
evolution which is obtained by solving the Boltzmann equation. 
The relic density can finally be expressed in terms of
the critical density
\begin{eqnarray}
\Omega_D^{} h^2 \simeq { 1. 07 \times 10^9~ {\rm GeV^{-1}} \over
M_{Pl}^{}} {x_f \over \sqrt{g_\ast}} {1 \over a+3 b/x_f }
\end{eqnarray}
where $h$ is the Hubble constant in units of $100 ~{\rm km} / {\rm
s\cdot Mpc}$, $M_{Pl}^{}= 1.22 \times 10^{19} ~{\rm GeV}$ is the
Planck mass and $g_\ast$ is the number of relativistic degrees of
freedom with mass less than $T_f^{}$. The freeze-out temperature
$x_f^{}$ can be estimated through the iterative solution of the
equation \cite{darkrep}
\begin{eqnarray}
x_f^{} = \ln \left[ c(c+2) \sqrt{45\over 8} {g \over 2\pi^3} {{
M_{Pl}^{}m_\psi^{}} \langle \sigma_{ann}v_{rel}\rangle \over
\sqrt{g_\ast x_f^{}} } \right]\simeq {\rm ln}{0.038 M_{Pl}m_\psi
(a+6b/x_f) \over \sqrt{g_\ast x_f}}
\end{eqnarray}
where $c$ is the constant of order one determined by matching the
late-time and early-time solutions and $g$ is the weak interaction
gauge coupling constant.

\begin{figure}[tb]
\begin{center}
\begin{tabular}{cc}
\includegraphics[scale=1,width=12cm]{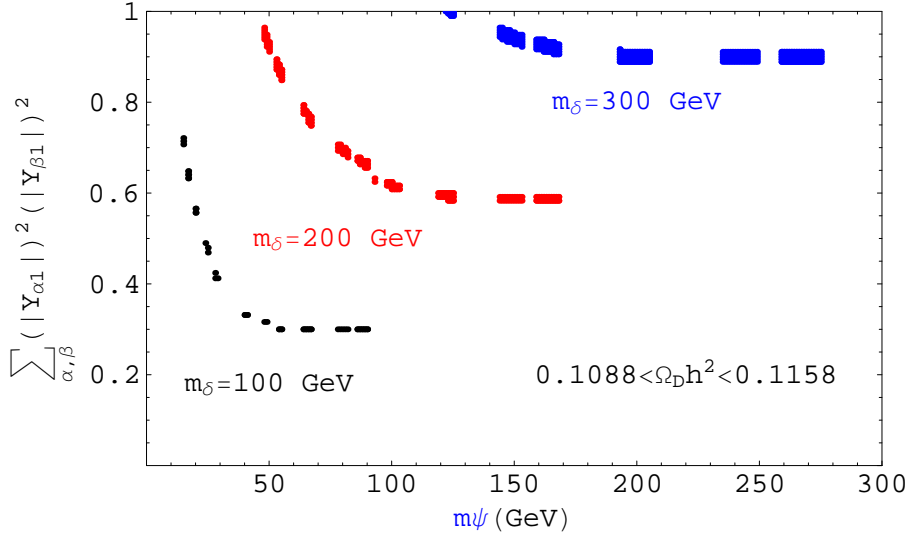}
\end{tabular}
\end{center}
\caption{$\sum_{\alpha\beta}|(Y_\psi)_{\alpha 1}|^2|(Y_\psi)_{\beta
1}|^2$ versus dark matter mass $m_\psi$ constrained by dark matter
relic density.} \label{density}
\end{figure}

In Fig.~\ref{density} we show $\sum_{\alpha \beta}^{}
|(Y_\psi)_{\alpha 1}^{}|^2 |(Y_\psi)_{\beta 1}^{}|^2 $ versus dark
matter mass $m_\psi^{}$ constrained by dark matter relic density
with different $\delta^\pm$ masses $m_\delta=100~{\rm GeV}, 200~{\rm
GeV}, 300~{\rm GeV}$. We find, to get correct relic density, Yukawa
coupling $Y_{\psi}^{}$ should be at order ${\cal O} (1)$ with a few
hundred GeV $m_\psi^{}$ and $m_\delta$.

\section{The test of Radiative Seesaw in B-L extension at the LHC}
In order to study the predictions for the lepton flavor correlation
and dark matter property with the radiative seesaw mechanism in
$U(1)_{B-L}$ extension, the ideal channel is production of new scalar $\delta$ and
fermion $\psi$ via the mediation of new gauge boson
$Z'$ in this theory.
\subsection{New Gauge Boson $Z'$ Decay}
In the limit where there is no mixing between the two Abelian
sectors of the minimal $B-L$ extension of the SM, the mass of the
new gauge boson $Z'$ is given as $M_{Z'}=2g_{BL}v_\Phi$, see
Eq.~(\ref{zblmass}). To satisfy the experimental lower bound,
$M_{Z'}/g_{BL}>5-10$ TeV~\cite{LEP}, it is sufficient to assume that
$v_\Phi>2.5-5$ TeV. There has been a lot of works on the heavy
neutral gauge bosons. For a recent review, see
Ref.~\cite{Langacker:2008yv}, and recent studies of $Z'$ at the
Tevatron and LHC~\cite{Petriello:2008zr}. For a recent study of the
phenomenological aspects of the $B-L$ model, see Ref.~\cite{Huitu}.
In Ref.~\cite{ourbl} heavy Majorana neutrino pair production via
$Z'$ is well-studied when $M_N<M_{Z'}/2$.

\begin{figure}[tb]
\begin{center}
\begin{tabular}{cc}
\includegraphics[scale=1,width=12cm]{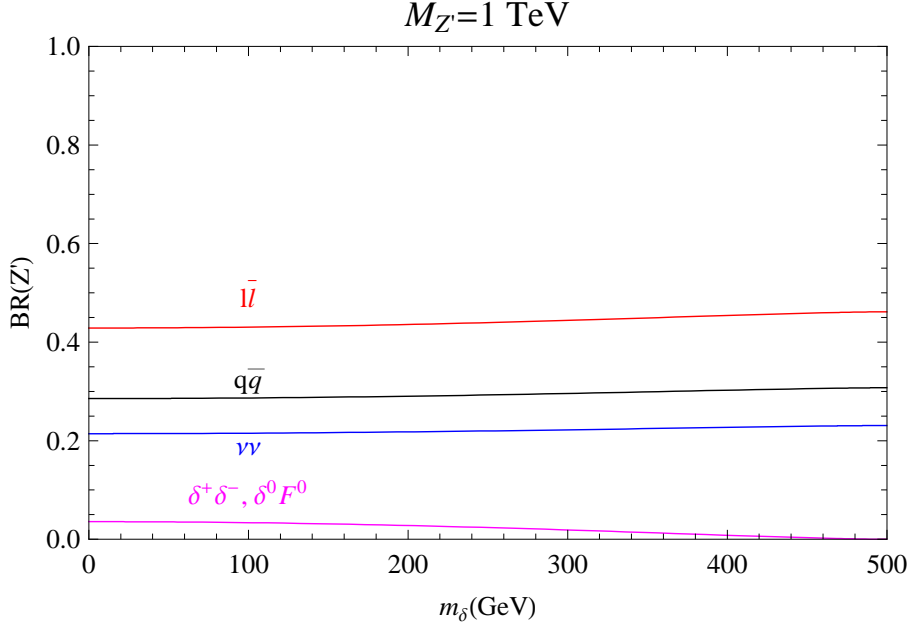}
\end{tabular}
\end{center}
\caption{The branching ratios of $Z'$ decay versus $m_\delta$ when
$M_{Z'}=1$ TeV.} \label{zpbr}
\end{figure}

Here, as indicated in earlier section, we pay attention to the
observable signatures in this model when the case $M_N>M_{Z'}/2$
happens and thus the pair production of heavy Majorana neutrinos
from on-shell $Z'$ is forbidden. Therefore, with the mass
hierarchy given in Eq.~(\ref{hierarchy}), the partial widths of $Z'$ decay
are
\begin{eqnarray}
\Gamma(Z'\to f\bar{f})&=&g_{BL}^2 {M_{Z'}\over 12\pi} C_f
(Q_{BL}^f)^2
\left(1+2{m_f^2\over M_{Z'}^2}\right)\ \beta_f , \\
\Gamma(Z'\to \sum_m \nu_m \nu_m)&=& 3 g_{BL}^2 {M_{Z'}\over
24\pi}C_\nu (Q_{BL}^\ell)^2,\\
\Gamma(Z'\to \delta^+\delta^-)&=&{g_{BL}^2\over 48\pi}M_{Z'}\beta_{\delta^\pm}^3
={g_{BL}^2\over 48\pi}M_{Z'}\beta_{\delta}^3,\\
\Gamma(Z'\to \delta^0F^0)&=&{g_{BL}^2\over
48\pi}M_{Z'}\left[1-2{m_{\delta^0}^2+m_{F^0}^2\over
M_{Z'}^2}+{(m_{\delta^0}^2-m_{F^0}^2)^2\over
M_{Z'}^4}\right]^{3/2}={g_{BL}^2\over 48\pi}M_{Z'}\beta_{\delta}^3
\end{eqnarray}
where $f=\ell,q$, the couplings $C_{\ell,\nu}=1,C_q=3$, and $\beta_i
= \sqrt{1- 4m_i^2/ M_{Z'}^2}$ is the speed of particle $i$. In
Fig.~\ref{zpbr} we plot the branching ratios of $Z'$ decay versus
$m_\delta$ when $M_{Z'}=1$ TeV. Notice the mass of new scalar $m_\delta$ is constrained by
dark matter relic density as several
hundred GeV. One can see, in low $m_\delta$ range, the $Z'$ decay branching fractions
take simple ratios for the final states
\begin{eqnarray}
\sum_{\ell}^{e,\mu,\tau}\ell^+\ell^-:\sum_q^{u\cdots t}q\bar{q}:
\sum_m^{1,2,3}\nu_m\nu_m:\delta^+\delta^-:\delta^0F^0=3:2:{3\over 2}:{1\over 4}:{1\over 4}.
\end{eqnarray}

\subsection{New Charged Scalar $\delta^\pm$ Decay}
As seen in previous section, the new scalar $\delta$ only couples to
the stable particle $\psi$ and SM leptons because of $B-L$ gauge
invariance. The leading decay channels for the new scalar include
$\delta^\pm\to \ell^\pm \psi_{1,2}$ and $\delta^0\to \nu(\bar{\nu})
\psi_{1,2}$. All relevant amplitudes are proportional to the Yukawa
coupling $Y_\psi$ which contributes to light neutrino mass
generation. Therefore the lepton-flavor contents of $\delta$ decay
will be different in each neutrino spectrum.

\begin{figure}[tb]
\begin{center}
\begin{tabular}{cc}
\includegraphics[scale=1,width=8cm]{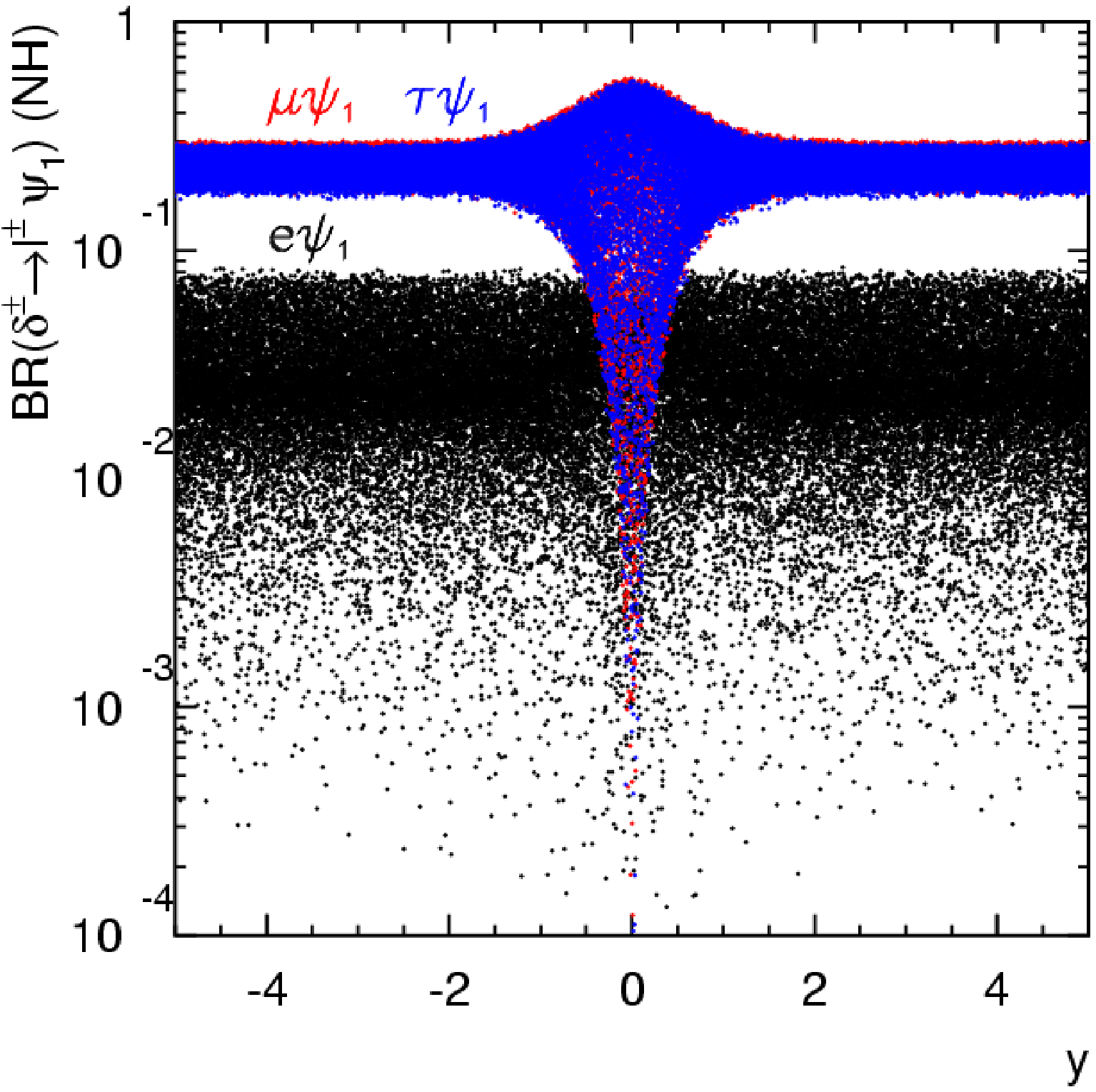}
\includegraphics[scale=1,width=8cm]{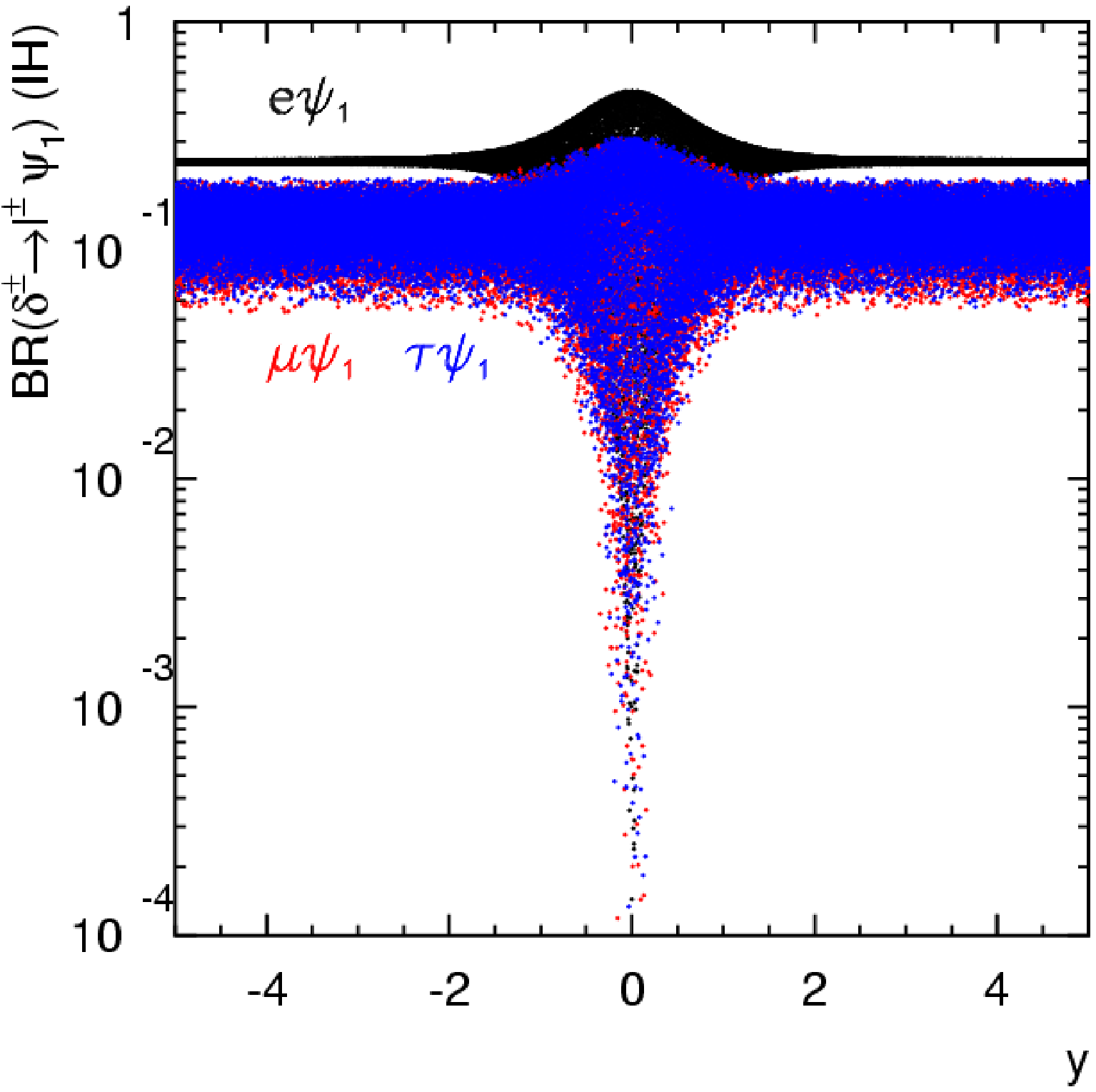}
\end{tabular}
\end{center}
\caption{The branching ratios of $\delta^\pm\to \ell^\pm \psi_1 \
(\ell=e,\mu,\tau)$ versus parameter $y$ in matrix $O$ for NH (left)
and IH (right), assuming vanishing Majorana phase, $0\leq x\leq
2\pi$.} \label{deltabrv}
\end{figure}

\begin{figure}[tb]
\begin{center}
\begin{tabular}{cc}
\includegraphics[scale=1,width=8cm]{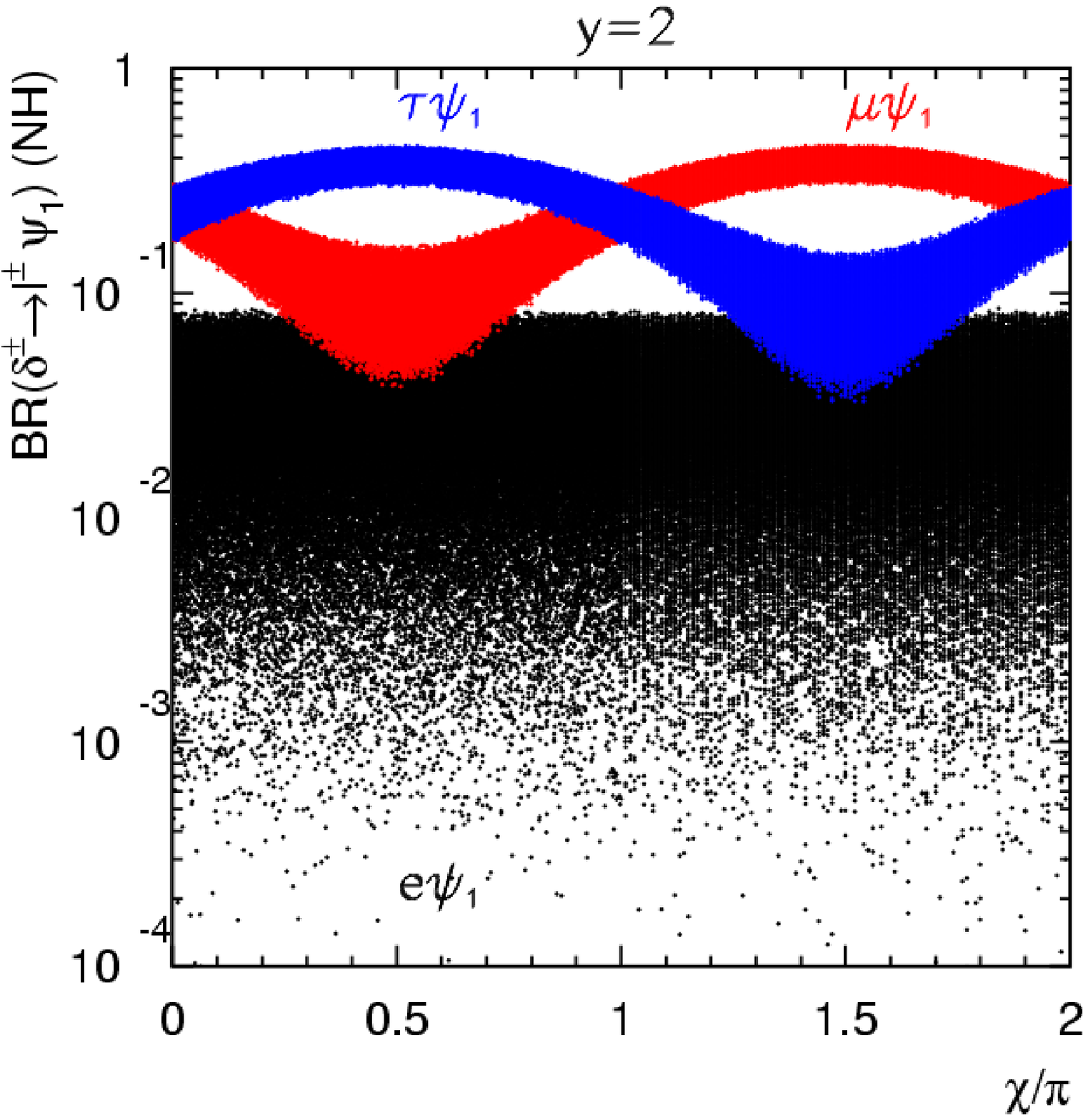}
\includegraphics[scale=1,width=8cm]{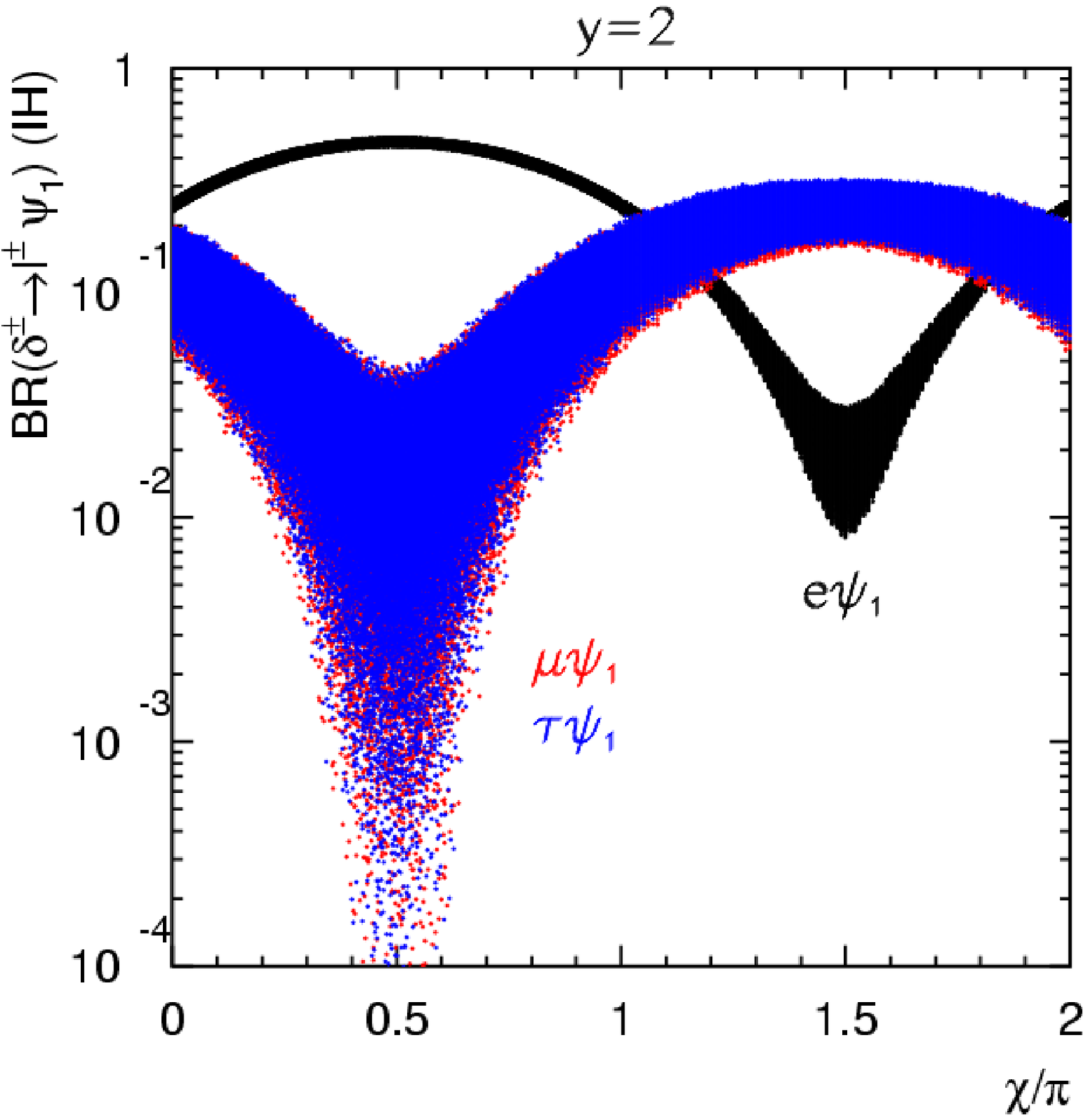}
\end{tabular}
\end{center}
\caption{The branching ratios of $\delta^\pm\to \ell^\pm \psi_1 \
(\ell=e,\mu,\tau)$ versus Majorana phase $\chi$ for NH (left) and IH
(right), when $y=2$ and $0\leq x\leq 2\pi$.} \label{deltabrph}
\end{figure}

The partial widths of $\delta^\pm$ decay are
\begin{eqnarray}
\Gamma(\delta^\pm\to \ell^\pm\psi_{1(2)})&=&{|(Y_\psi)_{\ell
1(2)}|^2\over 16\pi m_{\delta^\pm}}\lambda^{1/2}(1,{m_\ell^2\over
m_{\delta^\pm}^2},{m_{\psi_{1(2)}}^2\over
m_{\delta^\pm}^2})(m_{\delta^\pm}^2- m_\ell^2-m_{\psi_{1(2)}}^2)
\end{eqnarray}
where $\lambda(x,y,z)=x^2+y^2+z^2-2xy-2xz-2yz$. The dependence of
$\delta^\pm$ decay branching fractions $\delta^\pm\to \ell^\pm
\psi_1 \ (\ell=e,\mu,\tau)$ on parameter y is plotted in
Fig.~\ref{deltabrv}. The behaviors for $\psi_2$ final state are
almost the same. One can see that if SM charged lepton masses are
ignored, the decay branching fractions do not depend on both masses
of $\delta$ and $\psi_1$ entirely, and the parameter $y$ except for
range $|y|\lesssim 1$. The branching fractions can differ by one
order of magnitude in NH case $BR(e^\pm\psi_1)\ll
BR(\mu^\pm\psi_1),BR(\tau^\pm\psi_1)\approx 30\%$ and a few times in
the IH spectrum
$BR(\mu^\pm\psi_1),BR(\tau^\pm\psi_1)<BR(e^\pm\psi_1)\approx 25\%$.
Therefore one can distinguish neutrino mass spectra according to
different SM lepton flavors of dominant channels in final states. In
Fig.~\ref{deltabrph}, we show the dependence of $\delta^\pm$ decay
branching fractions on Majorana phase $\chi$ in NH and IH for $y =
2$. In NH the dominant channels swap from $\tau^\pm \psi_1$ when
$\chi\approx \pi/2$ to $\mu^\pm \psi_1$ when $\chi\approx 3\pi/2$ by
a few times. In IH the dominant channels swap from $e^\pm \psi_1$
when $\chi\approx \pi/2$ to $\mu^\pm \psi_1,\tau^\pm \psi_1$ when
$\chi\approx 3\pi/2$ by more than one order of magnitude. This
qualitative change can be made use of extracting the value of the
Majorana phase $\chi$ and parameter $y$. Moreover, it is important
to note that the curves of branching fractions corresponding to
Majorana phase translate parallelly by a phase $\pi$ for $-y$ case.

\subsection{Production and Mass Determination of Dark Matter $\psi$ and $\delta^\pm$ at the LHC}
\begin{figure}[tb]
\begin{center}
\begin{tabular}{cc}
\includegraphics[scale=1,width=8cm]{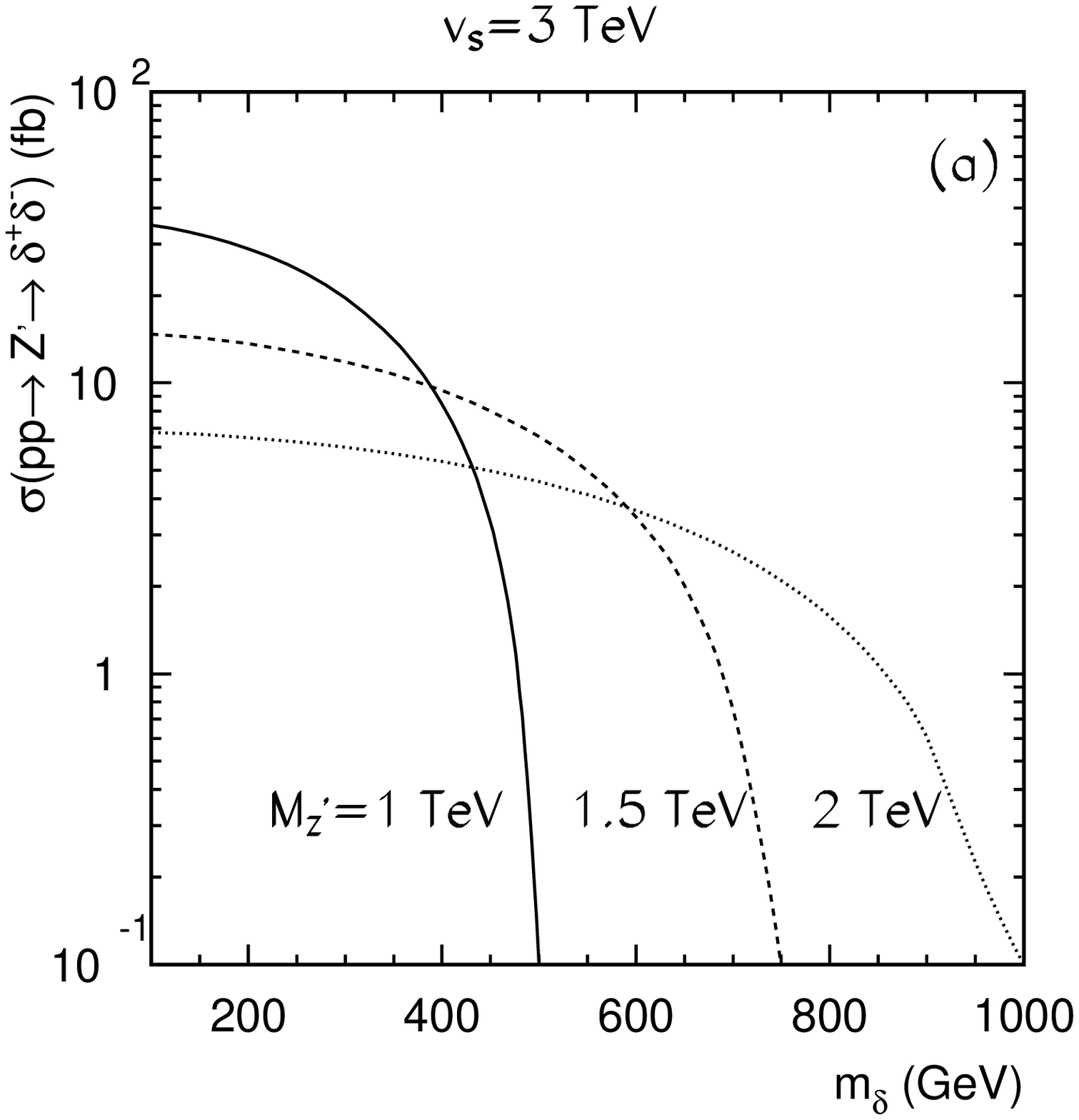}
\includegraphics[scale=1,width=8cm]{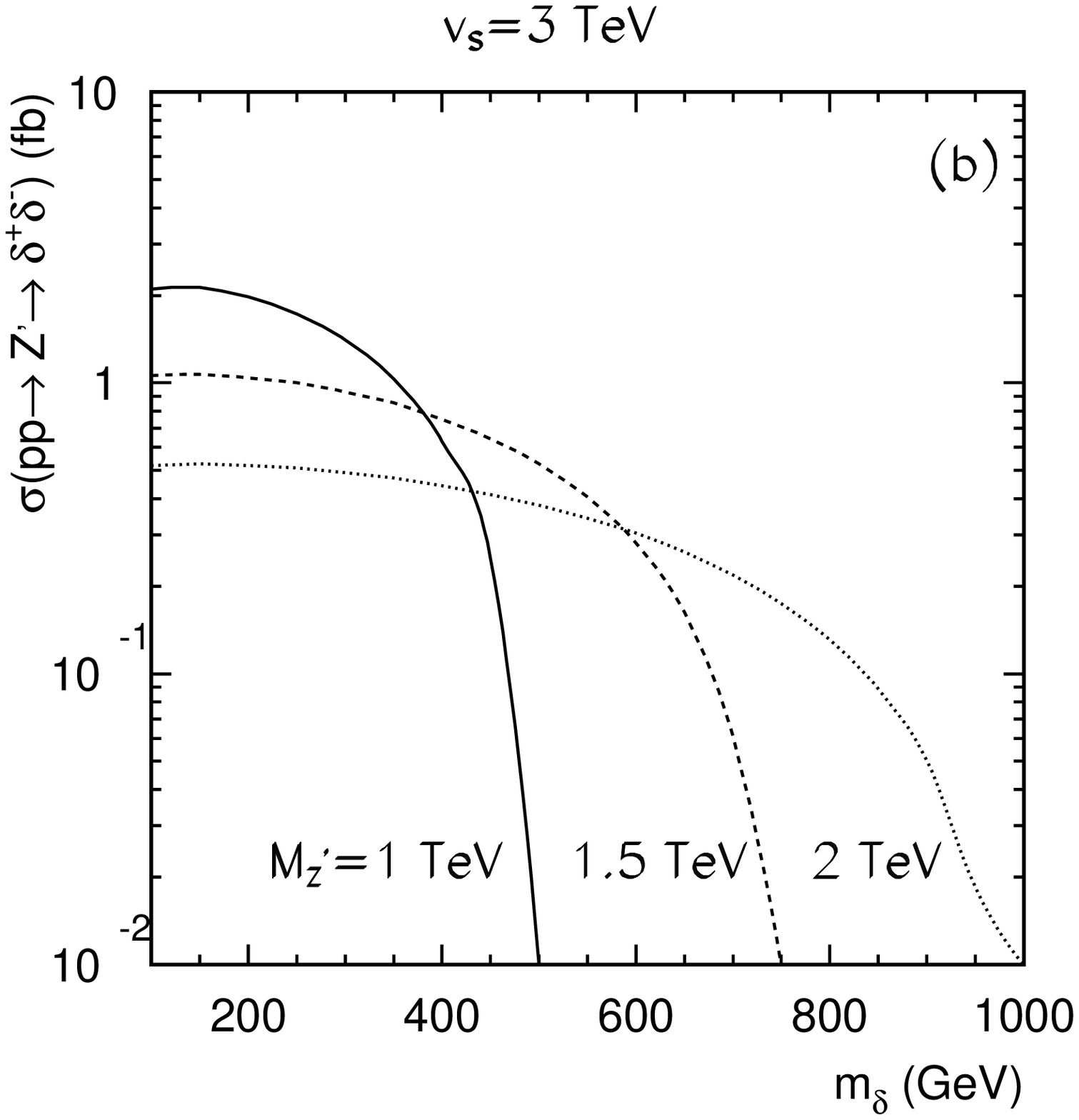}
\end{tabular}
\end{center}
\caption{Charged scalar $\delta^\pm$ pair production total cross
section at the LHC versus its mass $m_\delta$ (a) without any cuts
and (b) with all cuts and branching fraction of $\delta^\pm$ decay
$BR(\delta^\pm\to \ell^\pm \psi)=30\%$ taken from
Fig.~\ref{deltabrv}. The solid, dashed and dotted curves are for
$M_{Z'}=1,1.5,2~{\rm TeV}$ respectively, when $v_\Phi=3~{\rm TeV}$.}
\label{crosssection}
\end{figure}

Since in this model one has a dynamical mechanism for $B-L$
breaking, there is a typical pair production mechanism of
$\delta^\pm$ through $Z'$ with $\delta^\pm$ decay into SM charged
leptons and the stable particle $\psi$
\begin{eqnarray}
pp\to Z'\to \delta^+\delta^-\to \ell^+\ell^-\psi\psi .
\end{eqnarray}
In Fig.~\ref{crosssection} (a) we plot the total cross section of
charged scalar $\delta^\pm$ pair production at the LHC versus its
mass $m_\delta$. Considering the mass difference between
$\delta^\pm$ and $\psi$ can be large, see Fig.~\ref{density}, our
signal would be two hard opposite-sign leptons plus large missing
energy. The irreducible SM backgrounds are $Z(\to
\nu\bar{\nu})Z/\gamma^\ast(\to \ell^+\ell^-)$ and $W^+(\to
\ell^+\nu)W^-(\to \ell^-\bar{\nu})$. For our numerical analyzes, we
adopt the CTEQ6L1 parton distribution function~\cite{CTEQ} and
evaluate the SM backgrounds by using the automatic package
Madgraph~\cite{Madgraph}. We work in the parton-level, but simulate
the detector effects by the kinematical acceptance and employ the
Gaussian smearing for the electromagnetic energies~\cite{cms}
\begin{eqnarray}
{ \Delta E\over E} &=& {a_{cal} \over \sqrt{E/{\rm GeV}} } \oplus
b_{cal}, \quad a_{cal}=10\%,\  b_{cal}=0.7\% . \label{ecal}
\end{eqnarray}
We employ the following basic acceptance cuts for the event selection~\cite{cms}
\begin{eqnarray}
&&p_T(\ell)\geq 15~{\rm GeV}, \ |\eta(\ell)|<2.5, \ \Delta
R_{\ell\ell}\geq 0.4, \ \cancel{E}_T>30~{\rm GeV}
\end{eqnarray}
Because the two leptons can be very hard, we tighten up the
transverse momenta of them
\begin{eqnarray}
p_{T}(\ell)>120 \ \rm GeV.
\end{eqnarray}
This cut also helps for the background reduction significantly. The
total cross section with all the cuts being set is plotted in
Fig.~\ref{crosssection} (b), in which we take the branching fraction
of $\delta^\pm$ as $BR(\delta^\pm\to \ell^\pm \psi)=30\%$ from
Fig.~\ref{deltabrv}. For lower mass range of $m_\delta$ and $M_{Z'}$
one can have $\mathcal{O}(100)$ event number with $100$ fb$^{-1}$
integrated luminosity.

\begin{figure}[tb]
\begin{center}
\begin{tabular}{cc}
\includegraphics[scale=1,width=8cm]{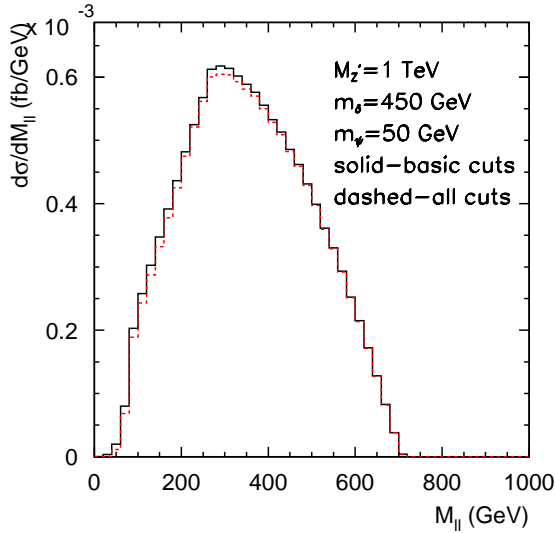}
\end{tabular}
\end{center}
\caption{Invariant mass distribution of two leptons for $pp\to Z'\to
\delta^+\delta^-\to \ell^+\ell^-\psi\psi$ production after basic
cuts and all cuts, with $M_{Z'}=1$ TeV, $m_\delta=450$ GeV and
$m_\psi=50$ GeV.} \label{mll}
\end{figure}

On the other hand, the masses of $\psi$ and $\delta^\pm$ can be
well-determined in this topology with the help of so-called cusp
kinematics~\cite{cusp}. Considering the charged leptons in final
states to be massless, due to the on-shell constraint for the
particle $\delta^\pm$ there is a cusp and end-point in invariant
mass distribution of the two leptons
\begin{eqnarray}
&&M_{\ell\ell}^{cusp}=m_\delta\left(1-{m_\psi^2\over
m_\delta^2}\right)e^{-\zeta}, \
M_{\ell\ell}^{max}=m_\delta\left(1-{m_\psi^2\over
m_\delta^2}\right)e^{\zeta}, \ {\rm cosh} \zeta={M_{Z'}\over
2m_\delta},
\end{eqnarray}
where $\zeta$ is the rapidity of $\delta^\pm$. Taking $M_{Z'}=1$
TeV, $m_\delta=450$ GeV and $m_\psi=50$ GeV as reference masses, we
plot $M_{\ell\ell}$ in Fig.~\ref{mll}. One can read the two
variables from this plot as $M_{\ell\ell}^{cusp}=280$ GeV and
$M_{\ell\ell}^{max}=708$ GeV. Once we know the mass of gauge boson
$Z'$ from purely leptonic final states $pp\to Z'\to \ell^+\ell^-$
future~\cite{zpll}, the masses of stable particle $\psi$ and its
parent particle $\delta^\pm$ can be exactly solved from the two
equations given above. The determined masses are $m_\delta=450.6$ GeV and
$m_\psi=49.4$ GeV in this case.

\section{Conclusion and Summary}

In this paper we have established a hybrid seesaw mechanism to
explain tiny neutrino masses and suggest a cold dark matter candidate. 
In this mechanism which is beyond the so-called Type I seesaw in the context of 
$B-L$ gauge
symmetry extension of the Standard Model, the contributions of a new scalar doublet 
and two new fermion singlets appear at one-loop level. We have studied in detail
the constraints on the model parameters from neutrino oscillation data, lepton flavor
violating processes and cosmological observation. We have also
explored the predictions on the decay branching ratios of the charged new scalar in each
neutrino mass spectrum and showed the most optimistic
scenarios where one could hope to distinguish the spectra using the
properties of the decays. The typical signatures related to the new
seesaw mechanism and dark matter candidate at the LHC are also
studied. We summarize our main results in the following
\begin{itemize}
\item A radiative seesaw mechanism can be added to the minimal Type I seesaw
with a local gauge symmetry $B-L$, a new scalar doublet and two new
fermion singlets $\psi$ at loop level.
\item The decays of new scalar doublet are related to neutrino
masses and mixings. We show the possibility to distinguish the
neutrino spectra. The branching fractions can differ by one order of
magnitude in NH case with $BR(\mu^\pm \psi),BR(\tau^\pm \psi)\gg
BR(e^\pm \psi)$, and a few times in the IH spectrum with $BR(e^\pm
\psi)>BR(\mu^\pm \psi),BR(\tau^\pm \psi)$ when the Majorana phase is
ignored in 3+2 mode.
\item Considering effects of the Majorana phase, in NH the dominant channels swap
from $\tau^\pm \psi_1$ when
$\chi\approx \pi/2$ to $\mu^\pm \psi_1$ when $\chi\approx 3\pi/2$ by
a few times. In IH the dominant channels swap from $e^\pm \psi_1$
when $\chi\approx \pi/2$ to $\mu^\pm \psi_1,\tau^\pm \psi_1$ when
$\chi\approx 3\pi/2$ by more than one order time of magnitude.
\item The lightest new particle $\psi$ in the loop is stable and can be cold dark
matter candidate because of $B-L$ gauge symmetry invariance.
Cosmological observation constrains the masses of the new scalar and
$\psi$ at a few hundred GeV.
\item Even when heavy Majorana neutrino pair production channel is not allowed
from $Z'$ gauge boson decay, in this framework, the new charged
scalar can be pair produced at the LHC essentially.
\item The masses of new scalar $\delta^\pm$ and missing particle $\psi$
can be well-determined in the production topology $pp\to Z'\to
\delta^+\delta^-\to \ell^+\ell^-\psi\psi$ in terms of properties of
visible SM charged leptons invariant mass distribution.
\end{itemize}
%%%%%%%%%%%%%%%%%%%%%%%%%%
\subsection*{Acknowledgment}
%%%%%%%%%%%%%%%%%%%%%%%%%%
We acknowledge T.~Han for providing his Fortran codes HANLIB for our
calculations. T.~L. would like to thank X.~G.~He for helpful
discussions. This work is supported in part by the DOE under grant No. DE-FG02-91ER40626 (T.L.).

%%%%%%%%%%%%%%%%%%%%%%%%%%%%%%%%%%%%%%%%%%%%%%%%%%%%%%%%%%%%%%%%%%%%%%%%%
\appendix

\section{Feynman Rules}
We summarize the Feynman rules for our model in Table.~\ref{rule}.

\begin{table}[tb]
\begin{center}
\begin{tabular}[t]{|c|c|c|}
  \hline
  Fields & Vertices & Couplings\\
  \hline
  $\delta^\pm$ & $Z'_\mu \delta^+(p_1)\delta^-(p_2)$ & $-ig_{BL}(p_1-p_2)_\mu$\\
               & $Z_\mu \delta^+(p_1)\delta^-(p_2)$ & $-ig_2{\cos(2\theta_W)\over 2\cos(\theta_W)}(p_1-p_2)_\mu$\\
               & $A_\mu \delta^+(p_1)\delta^-(p_2)$ & $-ie(p_1-p_2)_\mu$\\
  \hline
  $\delta^0,F^0$&$Z'_\mu \delta^0(p_1)F^0(p_2)$&$-g_{BL}(p_1-p_2)_\mu$\\
                &$Z_\mu \delta^0(p_1)F^0(p_2)$&${\sqrt{g_1^2+g_2^2}\over 2}(p_1-p_2)_\mu$\\
  \hline
  $Z'$ & $\bar{q}_iq_iZ'$ & $-iQ^q_{BL}g_{BL}\gamma^\mu$\\
       & $q_1=u, q_2=d$   & $Q^q_{BL}={1\over 3}$\\
  \hline
       & $\bar{\ell}\ell Z'$   & $-iQ^\ell_{BL}g_{BL}\gamma^\mu$\\
       & $\ell=e,\mu,\tau$ & $Q^\ell_{BL}=-1$\\
  \hline
         & $\overline{N_{m}}N_{m}Z'$ & $ig_{BL}\gamma^\mu {\gamma_5\over 2}$\\
         & $N=\nu_R+\nu_R^C$ & \\
  \hline
       & $\overline{\nu_{m}}\nu_{m}Z'$ & $ig_{BL}\gamma^\mu {-\gamma_5\over 2}$\\
       & $\nu=\nu_L+\nu_L^C$ & \\
  \hline
       & $\phi^0(p_1)\phi^0(p_2)Z'$ & $-i2g_{BL}(p_1-p_2)_\mu$\\
  \hline
  $\psi$ & $\bar{\nu}\delta^0\psi$ & $-iY_\psi{1\over \sqrt{2}} P_R$\\
  & $\psi=\psi_R+\psi_R^C$ & \\
  \hline
   & $\bar{\nu}F^0\psi$ & $-Y_\psi{1\over \sqrt{2}} P_R$\\
  \hline
    & $\bar{\ell}\delta^-\psi$ & $iY_\psi P_R$\\
  \hline
\end{tabular}
\end{center}
\caption{Feynman rules. The momenta are all assumed to be incoming.}
\label{rule}
\end{table}

\end{document}